\documentclass[12pt]{article}
\usepackage{graphicx}        
\begin{document}

{\bf \large Computer Simulations of Opinions}
\bigskip

{\bf \large and Their Reactions to Extreme Events}

\bigskip
Santo Fortunato$^1$ and Dietrich Stauffer$^2$
\bigskip

$^1$ Faculty of Physics, Bielefeld University, D-33615 Bielefeld, Germany
\texttt{fortunat@physik.uni-bielefeld.de}

\bigskip
$^2$ Laboratoire de Physique et M\'ecanique des Milieux H\'et\'erog\`ene, 
\'Ecole Sup\'erieure de Physique et de Chimie Industrielles de la Ville de 
Paris, 10 rue Vauquelin, 74231 Paris Cedex 05, France and (permanent address):
Institute for Theoretical Physics, Cologne University, 
D-50923 K\"oln, Euroland 
\texttt{stauffer@thp.uni-koeln.de} 

\bigskip
{\small We review the opinion dynamics in the computer models of 
Deffuant et al. (D), of Krause and Hegselmann (KH), and of Sznajd (S). 
All these models 
allow for consensus (one final opinion), polarization (two final opinions),
and fragmentation (more than two final opinions), depending on how tolerant
people are to different opinions. We then simulate the reactions of people to extreme
events, in that we modify the opinion of an individual and investigate how the
dynamics of a consensus model diffuses this perturbation among the other
members of a community. It often happens that the
original shock induced by the extreme event influences the opinion
of a big part of the society.}

\section{Introduction}
\label{sec:1}

Predicting extreme events is very important when we want to avoid the losses
due to earthquakes, floods, stock market crashes, etc. But it is not easy,
as reading the newspapers shows. It is much easier to claim afterwards that
one has an explanation for this event. A more scientific question is the
investigation of the opinions people have after an extreme event: Do they
now take objective risks more seriously than before? Do people tend to 
exaggerate the risks and prefer to drive long distances by car instead of 
airplane, shortly after a plane crash happened? How do these opinion changes
depend on the time which has elapsed since the event, or the geographical
distance? It is plausible that the more time has elapsed since the last 
catastrophe, the less serious is the risk taken by most people. Less clear
is the influence of geographical distance, e.g. if the probability to die of 
a terror attack in a far away country is compared with the ``customary'' risk 
to die there from a traffic accident. 

Geipel, H\"arta and Pohl \cite{Pohl} looked at the geography question in a 
region of Germany where $10^4$ years ago a vulcano erupted and left the
Laach lake. The closer the residents were to that lake, the more seriously
they took the risk. But also their general political orientation was 
correlated with their risk judgment. On the other hand, scientific annoucements
led to some newspaper reactions within Germany, independent of the distance,
but died down after a few months. Other examples are the reactions to nuclear
power plants and their accidents. Volker Jentsch (private communication) 
suggested to simulate such reactions to extreme events on computers; such 
simulations are only possible with a reasonable model of opinion dynamics.

A very recent application would be the influence of deadly tsunamis after
an earthquake on the opinion of people. Those who live on the affected
coasts after that extreme event of December 2004 will remember it
as a clear danger. Those who life further inwards on the land, away from 
the coast, know that tsunamis do not reach them, but they still have learned
from the news about the thousands of people killed. Will they judge the danger
as higher or as lower than those on the affected coast line? And what about
those who live on the coast of a different ocean, where such events are 
also possible but happened long ago? This example shows how the influence
of an extreme event on the opinion of the people can depend on the distances 
in time and space. This is the question we want to simulate here in generic
models. 

It would not be desirable to invent a new opinion dynamic model just for the 
purpose to study reactions to extreme events. Instead, it would be nice if
one would have one generally accepted and well tested model, which then could
be applied to extreme events. No such consensus is evident from the literature.
We thus concentrate here on three models, D, KH and S (Deffuant et al.
\cite{Deffuant}, Krause and Hegselmann \cite{Hegselmann} and Sznajd 
\cite{Sznajd}) which are currently used a lot to simulate opinion dynamics; we 
ignore the older voter models \cite{voter} or those of Axelrod \cite{Axelrod},
of Galam \cite{Galam} and of Wu and Huberman \cite{Huberman}, to mention just 
some examples. We will not claim that as a result of these simulations one
may predict public reaction; we merely claim that simulations like these could
be a useful starting point in this research field. 

Of course, one may question in general whether human beings can be simulated
on computers where only a few numbers describe the whole person. More than
two millenia ago, the Greek philosopher Empedokles already paved the way to 
these computer simulation by stating (according to J. Mimkes), that 
some people are like wine and water, mixing easily, while others are like 
oil and water, refusing to mix. Thus he reduced the complexity of human 
opinions to two choices, like hydrophilic or hydrophobic in chemistry, 
spin up or spin down in physics, 0 or 1 in computer science. And in today's 
developed countries, we take regular polls on whether people like their
government, allowing only a few choices like: very much, yes, neutral,
no, or not at all. Simplifying mother Nature thus was not started by us, 
is common also in sociology, and has been quite successful in physics. 

\section{General Opinion Dynamics}
\label{sec:2}

In this section we review the dynamics of models D of Deffuant et al, KH of
Krause and Hegselmann, and S of Sznajd \cite{Deffuant,Hegselmann,Sznajd}. Their
results are quite similar but they differ in their rules on how the opinions 
are changed. An earlier review of these models was given in \cite{losalamos} 
with emphasis on the Sznajd model. In that model two people who agree in their 
opinions convince suitable neighbours to adopt this opinion. In model D, each person 
selects a suitable partner and the two opinions get closer to each other. 
For KH, each person looks at all suitable partners and takes their average 
opinion. ``Suitable'' means that the original opinions are not too far from 
each other. 

\subsection{Deffuant et al}
\label{sec:3}

\begin{figure}
\centering
\includegraphics[angle=-90, scale=0.45]{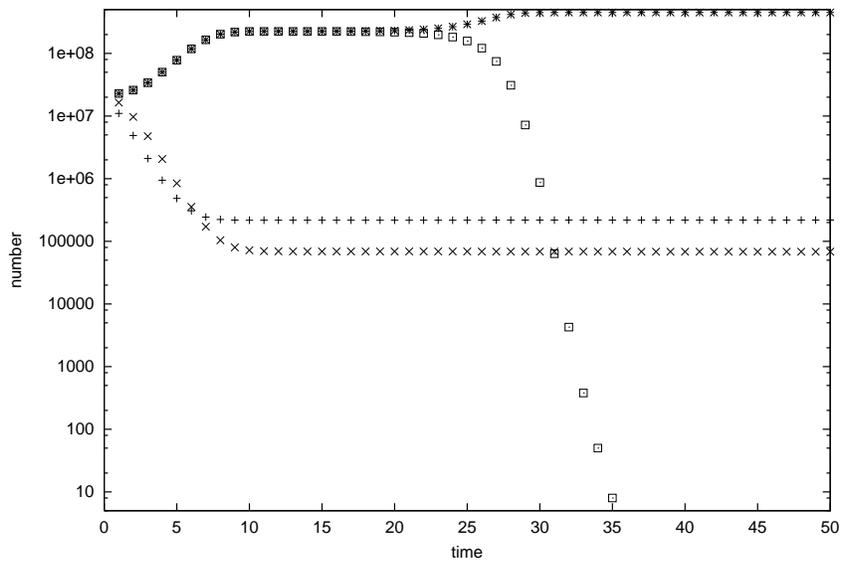}
\caption{Standard D model, 450 millions agents, $\epsilon = 0.4, \; \mu = 0.3$,
opinions divided in 20 intervals. Shown are intervals 1 (+), 2 (x), 10 (stars), 
and 11 (squares).} 
\label{fig:1}      
\end{figure}

In model D \cite{Deffuant} all $N$ agents have an opinion $O$ which can vary 
continuously between zero and one. Each agent selects randomly one of the 
other agents and checks first if an exchange of opinions makes sense. If the two 
opinions differ by more than $\epsilon \; (0 < \epsilon < 1)$, the two refuse
to discuss and no opinion is changed; otherwise each opinion moves partly in 
the direction of the other, by an amount $\mu \Delta O$, where $\Delta O$ is 
the opinion difference and $\mu$ the convergence parameter
($0<\mu<1/2$). The parameter $\epsilon$ is called confidence bound or confidence
interval.
For $\epsilon > 1/2$ all opinions converge towards 
a centrist one, while for $\epsilon < 1/2$ separate opinions survive; the 
number of surviving opinions in the latter case varies as $1/\epsilon$. Besides
simulations, also analytical approximations were made \cite{Redner} which 
agree well with the simulations.

Fig. \ref{fig:1} shows a consensus formation with the number of simulated people close to 
that in the European Union, 450 millions, and $\epsilon = 0.4,\; \mu = 0.3$. To
plot the results, the opinions were binned into 20 intervals. We show 
intervals 1, 2, 10 and 11 only. Initially, the numbers of opinions were the same
in all intervals; soon two centrist opinions dominate until finally one of them 
eats up the other. Independent of this power struggle, some extremist opinions
survive in the intervals close to zero and close to one. These extremist wings \cite{weis0}
are a general property for $\epsilon < 1/2$ but are not the theme of this 
``extreme'' book. 

Various variants of this standard version were published. 
It is numerically easier to look at integer opinions $O = 1,2,3 \dots, Q$ 
instead of continuously varying $O$; a precursor of such work is 
Galam and Moscovici \cite{mosco} where both discrete opinions (0, 1) 
and opinions in between were allowed. If the opinions $O = 1,2,\dots Q$ 
are integers, one can determine unambiguously if two opinions agree or differ.
The above expression for $\mu\Delta O$ then needs to be rounded to an integer.
If two opinions differ only by one unit, one randomly selected opinion is
replaced by the other one, whereas this other opinion remains unchanged.

The idea of everybody
talking to everybody with the same probability is perhaps realistic for
scientific exchanges via the internet, but, in politics, discussions on city
affairs are usually restricted to the residents of that city and do not
extend over the whole world. Putting agents onto a square lattice 
\cite{Deffuant} with interactions only between lattice neighbours 
\cite{Schelling} is one possibility. In recent years, small-world networks and 
scale-free networks \cite{Albert} were simulated intensively as models 
for social networks. In the standard version of the Barab{\'a}si-Albert model, 
the most popular model of scale-free networks,
one starts with a small number $m$ of agents all connected to each other. Then,
one by one, more members are added to the population. Each new member  
selects randomly $m$ previous members as neighbours such that the probability
of selecting one specific agent is proportional to the number of neighbours 
this agent had before. In this way, the well connected people get even more
connections, and the probability of one agent to have been selected as 
neighbour by $k$ later members is proportional to $1/k^3$. (In contrast, on the
square lattice and on the Bethe lattice, every agent has the same number of
neighbours, and for random graphs the number of neighbours fluctuates slightly
but its distribution has a narrow peak.) In opinion dynamics, only network
neighbours can influence each other. 
 
Putting Deffuant agents \cite{hmo}
onto this Barab{\'a}si-Albert network, with continuous
opinions, again for large confidence intervals $\epsilon$ a complete consensus
is found whereas for small $\epsilon$ the number of different surviving 
opinions varies roughly as $1/\epsilon$. An opinion cluster is a set of agents 
sharing in final equilibrium the same opinion, independent of whether these
agents are connected as neighbours or separated. Varying the total number $N$ 
of agents one finds that the number of small opinion clusters with 1, 2, 3,
... agents is proportional to $N$, while the number of large opinion clusters
comprising an appreciable fraction of the whole network is of order unity and
independent of $N$. This result reminds us of the cluster size distribution
for percolation \cite{book} above the threshold: There is one infinite cluster 
covering a finite fraction of the whole lattice, coexisting with many finite
clusters whose number is proportional to the lattice size. One may compare this
distribution of opinions with a dictatorship: The imposed official opinion coexists
with a clandestine opposition fragmented into many groups. 

This scale-free network can be studied in a complicated and a simple way: In
the complicated way, if a new agent Alice selects a previous agent Bob as 
neighbour of Alice, then Alice is also neighbour of Bob, like in mutual 
friendships. This is the undirected case. The directed case is the simpler way:
Bob is a neighbour for Alice but Alice is not a neighbour for Bob; this 
situation corresponds more to political leadership: the party head does not
even know all party members, but all party members know the head. Apart from 
simplifying the programming, the directed case seems to have the same 
properties as the undirected one \cite{hmo}. 

Also changing from continuous to discrete opinions $O = 1,2, \dots, Q$ does 
not change the results much but it simplifies the simulation \cite{deffdis},
particularly when only people differing by one opinion unit discuss with
each other (corresponding to $\epsilon \sim 1/Q$). Again the number of opinion 
clusters varies proportional to $N$ for $N \rightarrow \infty$ at fixed 
$N/Q$. A consensus is reached for $Q=2$, but not for $Q > 2$. A  scaling law 
gives the total number of final opinions as being equal to $N$ multiplied by a 
scaling function of $N/Q$. This law has two simple limits: For $Q \gg N$ there 
are so many opinions per person that each agent has its own
opinion, separate from the opinions of other agents by more than one unit: no 
discussion, nobody changes opinion, $N$ clusters of size unity. In the 
opposite case $Q \ll N$, all opinions have lots of followers and thus most
of them survive up to the end. These simple limits remain valid also if 
people differing by up to $\ell$ opinion units (instead of $\ell = 1$ only) influence
each other; a consensus is then formed if $\ell/Q$ (which now plays the role
of the above $\epsilon$) is larger than 1/2. (The more general scaling law
for arbitrary $N/Q$ now becomes invalid). This threshold of $\epsilon 
= 1/2$, which has emerged so often in the previous examples, is 
supposed to be a universal feature of the Deffuant dynamics, as long as the
symmetry of the opinion spectrum with respect to 
the inversion right $\leftrightarrow$ left is not violated
\cite{Fortunatoeps}. The symmetry means that the opinions $O$ and $1-O$ ($Q-O$
for integer opinions)
are equivalent and can be exchanged at any stage of the dynamics without
changing the corresponding configuration. In this way, the histogram 
of the opinions is at any time symmetric with respect to the central opinion
$1/2$ ($Q/2$ for integer opinions).
If we instead let $O$ and $1-O$ ($Q-O$) play different roles the
threshold will in general be different. 
As a matter of fact, in \cite{Assmann}
one introduced such an asymmetry in that the "convincing power", expressed by
the parameter $\mu$, is no longer the same for all agents but it depends on the 
opinion of the agent. More precisely, $\mu$ increases with the opinion of the
individual, and this implies that those agents with low values of $O$ are less convincing
than those with high values of $O$. In this case the opinion distribution is 
no longer symmetric with respect to $O=1/2$ ($Q/2$) and 
the consensus threshold is larger than $1/2$.  

In all this work, first the scale-free network was constructed, and then
the opinion dynamics studied on the fixed network. Not much is changed if
opinion dynamics takes place simultaneously with network growth \cite{SousaD},
in agreement with Ising and Sznajd models \cite{Bonnekoh}.  

\subsection{Krause-Hegselmann}
\label{sec:4}

The KH model \cite{Hegselmann} was simulated less since only small systems
seemed possible to be studied. Only recently, for
discrete opinions, an efficient algorithm was found to study millions of agents
\cite{FortunatoKH}, compared with at most 300,000 for continuous opinions
\cite{cise}. Again we have opinions $O$ continuous between zero and one, or
discrete $O = 1, 2, \dots Q$. At every iteration, every agent looks at all 
other agents, and averages over the opinions of those which differ by not more
than $\epsilon$ (continuous opinions) or $\ell$ (discrete opinions) from its
own opinion. Then it adopts that average opinion as its own. As in the D model,
also the KH model shows a complete consensus above some threshold and
many 
different opinions in the final configuration if $\epsilon$ is very small.
However, in this case, there are two possible values for the threshold \cite{sanHK},
depending on how many neighbours an agent has on average: if this number of
neighbours, or average degree,
grows with the number of agents of the community,  
there is consensus for $\epsilon>\epsilon_0$, where $\epsilon_0\sim\,0.2$; if instead the average degree
remains finite when the population diverges, the consensus threshold is
$1/2$ as in the D model. 
Various ways of opinion averaging  
were investigated \cite{Hegs}. Hegselmann and Krause \cite{Hegselmann} also 
simulated asymmetric $\epsilon$ choices, which may depend on the currently 
held opinion.

\begin{figure}
\centering
\includegraphics[angle=-90, scale=0.45]{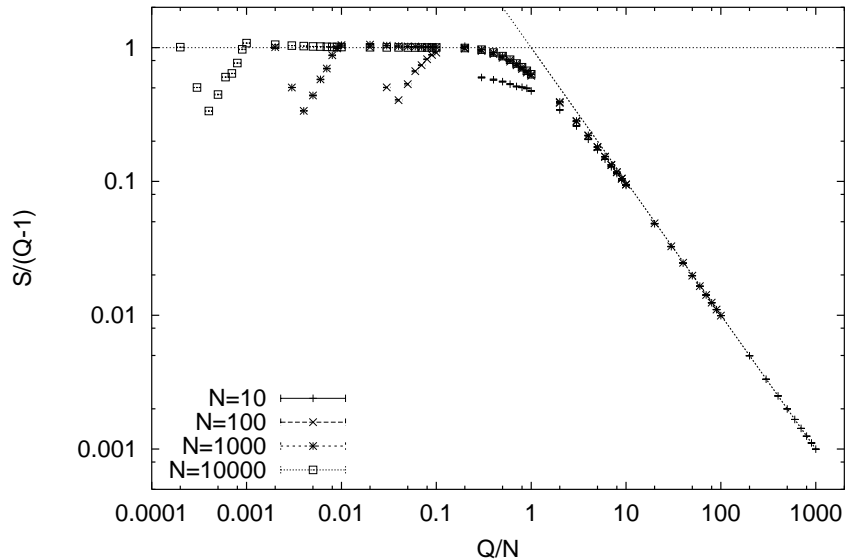}
\caption{Scaling law for the number $S$ of surviving opinions in the
discrete KH model, from \cite{FortunatoKH}. For the D model the figure
looks similar \cite{deffdis} except that the downward deviations at the 
left end of the data sets are weaker.}
\label{fig:2}
\end{figure}

Fig. \ref{fig:2} shows that the same scaling law as for the discrete D model also
holds for the discrete KH model \cite{FortunatoKH} on a scale-free 
Barab{\'a}si-Albert network. For the usual version of the model, in which all individuals
talk to each other, but with
discrete opinions and discussions only between agents differing by one opinion
unit, up to $Q = 7$ a consensus is reached while for $Q > 7$ several
opinions remain. (The role of well-connected leaders in a similar opinion
model on a Barab{\'a}si-Albert network was studied in \cite{Sun}.)

As we mentioned above, by using discrete opinions it is possible to 
speed up the algorithm compared to the continuous case.
The implementation of an algorithm for KH with discrete opinions must be probabilistic,
because the value of the average opinion of compatible neighbours of an agent
must necessarily be rounded to an integer and this would make the dynamics
trivial, as in most cases the agent would keep its own opinion. 
We start with a community where everybody talks to everybody else, 
opinions $O=1,2, .., Q$ and a
confidence bound $\ell$. 
After assigning at random opinions to the agents in the initial configuration,
we calculate the histogram $n_O$ of the opinion distribution, by counting how
many agents have opinion $O$, for any $O=1, 2, ..., Q$.
Suppose we want
to update the status of agent $i$, which 
has opinion $k$. The agents which are compatible with $i$ are all agents 
with opinion $\overline{k}=k-\ell, k-\ell+1, ..., k, k+1, ..., k+\ell-1, k+\ell$.
Let $n_{k\ell}=n_{k-\ell}+n_{k-\ell+1}+...+n_{k+\ell-1}+n_{k+\ell}$ 
be the total number of compatible agents.
Then we say that agent $i$ takes opinion $\overline{k}$ with the probability 
$p_{\overline{k}}=n_{\overline{k}}/n_{k\ell}$,
which just amounts to choosing at random one of the agents
compatible with $i$
and taking its opinion.
Let $k_f$ be the new opinion of
agent $i$.
We simply need to withdraw one agent from the
original channel $k$ and add it to the channel $k_f$ to have the new 
opinion histogram of the system, and we can pass to the next update. Notice that in this way
the time required for a sweep over the whole population goes like $(2\ell+1)N$,
where $N$ is as usual the total number of agents and $2\ell+1$ the number of
compatible opinions. 
In the original algorithm with continuous opinions, instead, the time to complete an
iteration goes as $N^2$,
because to update the state of any agent one needs to make a sweep over the whole
population to look for compatible individuals and calculate the 
average of their opinions. The gain in speed of the algorithm with discrete opinions is
then remarkable, especially when $\ell{\ll}N$.

We have seen that 
the presence of the second factor $N$ in the expression of the 
iteration time for the continuous model is exclusively due to the fact that we
consider a community where every agent communicates with all others.
If one instead considers social topologies where each agent interacts on
average with just a few individuals, like a lattice, the iteration time will  
grow only linearly with $N$, and the algorithm will compete in speed with that 
of D. As a matter of fact, in many such cases the KH algorithm 
is much faster than the D algorithm.

\subsection{Sznajd}
\label{sec:5}

The S model \cite{Sznajd} is the most often studied model, and the literature
up to mid-2003 was reviewed in \cite{losalamos}. Thus we concentrate here on
the more recent literature. 

The most widespread version uses a square lattice with two opinions, $O = \pm 1$. 
If the two opinions in a randomly selected neighbour pair agree, then these
two agents convince their six lattice neighbours of this opinion; otherwise
none of the eight opinions changes. If initially less than half of the opinions
have the value 1, at the end a consensus is reached with no agent having
opinion 1; if initially the 1's have the majority, at the end everybody follows
their opinion. Thus a phase transition is observed, which is the sharper the
larger the lattice is. The growth of nearly homogeneous domains of $-1$'s and 1's 
is very similar to spinodal decomposition of spin 1/2 Ising magnets. 

\begin{figure}
\centering
\includegraphics[angle=-90, scale=0.45]{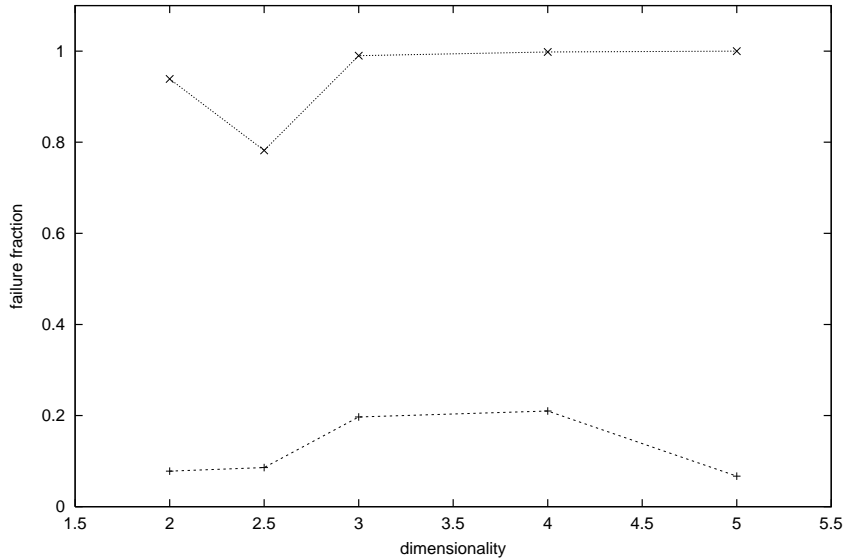}
\caption{Variation with dimensionality of the probability not to reach a
complete consensus. $d=2.5$ represents the triangular lattice. The upper data
refer to four opinions, the lower ones to three opinions, in small lattices:
$19^2$, $7^3$, $5^4$, $5^5$.
For larger lattices, the failures for three opinions vanish.
Opinion $O$ can only convince opinions $O \pm 1$.
}
\label{fig:3}
\end{figure}

With $Q>2$ possible opinions ($O=1, 2, ..., Q$), always a consensus is found except if 
only people with a neighboring opinion $O \pm 1$ can be convinced by the
central pair of opinion $O$; then a consensus is usually possible for
$Q \le 3$ but not for $Q \ge 4$ in a variety of lattice types and dimensions,
see Fig. \ref{fig:3} (from \cite{losalamos}).

The greatest success of the S model is the simulation of political election
results: The number of candidates receiving $v$ votes each varies roughly
as $1/v$ with systematic downward deviations for large and small $v$. This
was obtained on both a Barab{\'a}si-Albert \cite{bern} and a pseudo-fractal model
\cite{Gonzalez}. Of course, such simulations only give averages, not the
winner in one specific election, just like physics gives the air pressure
as a function of density and temperature, but not the position of one specific
air atom one minute from now. 

Schulze \cite{Schulze} simulated a multilayer S model, where the layer number
corresponds to the biological age of the people; the results were similar as 
for the single-layer S model. More interesting was his combination of global and
local interactions on the square lattice: two people of arbitrary distance who 
agree in their opinions convince their nearest neighbours of this opinion. 
Similarly to the mean field theory of Slanina and Lavicka \cite{Slanina}, the
times needed to reach consensus are distributed exponentially and are quite 
small. Therefore up to $10^9$ agents could be simulated. The width of the phase
transition (for $Q = 2$, as a function of initial concentration) vanishes 
reciprocally to the linear lattice dimension \cite{Schulze}.

If the neighbours do not always follow the opinion of the central pair, but
do so only with some probability \cite{Sznajd}, one may describe this 
probability through some social temperature $T$: The higher the temperature is,
the higher is the probability to change opinion \cite{He}. Then $T = 0$ means
nobody changes opinion, and $T = \infty$ means everybody follows the S rule.
Alternatively, one may also assume that some people permanently stick with
their opinion \cite{He,Schneider}. In this way, a more democratic society
is modeled even for $Q = 2$ such that not everybody ends up with the same
opinion. 

In an S model with continuous opinions and confidence bound $\epsilon$ 
similar to the D and KH models, always a consensus was found independently of 
$\epsilon$ \cite{FortunatoS}.

\section{Damage Spreading}
\label{sec:6}

How is it possible to describe the reaction of people to extreme events in
quantitative terms? From the previous discussion we have learnt that opinions
can be treated as numbers, integer or real. A change of opinion of an arbitrary
agent $i$ is thus simply the difference between the new opinion and the old one.
During the dynamical evolution, as we have seen above, opinions change, due to the
influence of the people on their acquaintances. This is, however, the "normal"
dynamics within a community. What we would like to investigate is instead 
how much a sudden perturbation ("extreme event") would alter the 
opinion variables of the agents of the system. 
The concept of perturbation need not be exactly
defined: for us it is whatever causes opinion changes in one or a 
few\footnote{Here "a few" means that the agents represent a negligible fraction 
of the total population, which vanishes in the limit of infinitely many agents.} agents of
the system. We have in mind localized events, like strikes, accidents, decisions
involving small areas, etc. We assume that people shape their own opinions only 
through the interactions with their acquaintances, without considering the
influence of external opinion-affecting sources like the mass media, which 
act at once on the whole population.

In order to evaluate the effect of a perturbation on the public opinion it is
necessary to know the opinion distribution of the agents when nothing anomalous
takes place ("normal state"), and compare it with the distribution determined  
after the occurrence of an extreme event. From the comparison between these two 
replicas of the system we can evaluate, among other things, 
the so-called Hamming distance, i. e. how many agents have changed their mind, and 
how the influence of the perturbation spread as a function of time and distance
from the place where the extreme event occurred.

This kind of comparative analysis is by no means new in science, and it is commonly
adopted to investigate a large class of phenomena, the so-called 
{\it damage spreading} processes. Damage spreading (DS) was originally introduced 
in biology by Stuart Kauffman \cite{kauf}, who wanted to estimate quantitatively
the reaction of gene regulatory networks to external disturbances ("catastrophic
mutations"). 
In physics, the first investigations focused on 
the Ising model \cite{stasta}. Here one starts 
from some arbitrary configuration of spins and creates a replica by flipping 
one or more spins; after that one lets both configurations evolve towards equilibrium
according to the chosen dynamics 
under the same thermal noise (i.e. identical sequences of random numbers).
It turns out that there is a temperature
$T_d$, near the Curie point, which separates a phase where the
damage heals from a phase in which the perturbation 
extends to a finite fraction of the spins of the system. 

The simplest thing one can do is just to follow the same procedure 
for opinion dynamics models. 
The perturbation consists in changing 
the opinion variable of an arbitrarily selected agent in the initial
configuration. After that, the chosen opinion dynamics applies for the two replicas. 
Preliminary studies in this direction already exist, and they deal with the   
Sznajd model on the square lattice. 
In \cite{Bernardes} one adopted a modified version of Sznajd where
the four agents of a plaquette convince all their neighbours
if they happen to share the same opinion; here the perturbed configuration is 
obtained by changing the opinion of all agents which lie on a line of the
lattice.
In \cite{Roehner} the shock consists in
the sudden change of opinions of some finite fraction $g$ of the whole
population and the time 
evolution of the number of perturbed agents is studied as a function of $g$.
More importantly, the authors of the latter paper show that in 
several cases critical shocks in social sciences can be used as probes to test
the cohesion of society. This recalls the strategy of natural sciences: 
if we hit an iron bar with a hammer, from the velocity of the sound in the bar
we are able to derive its density. 
In section \ref{sec:8} we will present new results
on damage spreading for the Sznajd opinion dynamics 
\cite{Klietsch}.
Here we focus on the other two consensus models, D and KH.
We shall first analyze the models for real-valued opinions,
then we will pass to integer opinions.
In all our simulations we defined the amount of damage as the number of 
agents differing in their opinions in an agent-to-agent comparison of the 
two replicas; we ignored the amount by which they differ.

An important issue is the choice of a suitable social topology.
A bidimensional lattice lends itself to 
a geographical description of the damage spreading process:
we can assume that the sites represent the position in space of the agents, and
that the "acquaintances" of an agent be its spatial neighbours. In this way
the lattice would map the distribution of people in some geographic area and
the distances between pairs of agents on the lattice can be associated to 
physical distances between individuals. On the other hand, the regular structure
of the lattice and the prescription of nearest-neighbour friendship 
endow the system with features which never occur in real communities.
In fact, on the lattice each agent has the
same number of friends and people who are
geographically far from each other are never friends. These unrealistic 
features can be removed by adopting a different kind of graph to describe the social
relationships between the agents. A Barab{\'a}si-Albert (BA) 
network \cite{BA} could be a good candidate:  
it is a non-regular graph 
where the number of acquaintances of an agent varies
within a wide spectrum of
values, with a few individuals having many friends whereas most 
people have just a few.
On the other hand the BA network is 
a structure with a high degree of
randomness and can hardly be embedded in an Euclidean bidimensional
surface, so a geographical characterization of the damage propagation  
would be impossible. In our opinion the ideal solution would be a graph 
which includes both the regular structure of the lattice and the disorder of a random 
graph. 
A possibility could be a lattice topology where 
the connection probability between the agents decays with some
negative power of the Euclidean distance, being unity for nearest neighbours.
In what follows we shall however 
consider only the square lattice and the BA network.

\subsection{Continuous Opinions}
\label{sec:7}

If opinions are real numbers, we need a criterion 
to state when the opinion of an agent
is the same in both replicas or different due to the initial perturbation. 
Since we used $64$-bit real numbers, we decided
that two opinions are the same if they differ
by less than $10^{-9}$. In order to determine with some precision 
the fraction of agents which changed their opinions, it is necessary to 
repeat the damage spreading analysis many times, by starting every time from a
new initial configuration without changing the set of parameters which constrain
the action of the dynamics: the final result is then 
calculated by averaging over all samples. 
In most our simulations we have collected $1000$ 
samples, in a few cases we enhanced the statistics up to $10000$.

For KH with continuous opinions a detailed damage spreading
analysis has recently been performed \cite{Fortunatodam}, for 
the case in which the agents sit on the sites of a BA network.
The dynamics of the KH model is fixed by a single parameter, the confidence
bound $\epsilon$, which plays the role of temperature in the Ising model.
Like in the Ising model, it is interesting to analyze
the damage propagation as a function of the control parameter
$\epsilon$; it turns out that there are three phases in the $\epsilon$-space,
corresponding to zero, partial and total damage, respectively.
The existence of a phase in which the initial perturbation manages to affect the 
state (here the opinions) of all agents is new for damage spreading processes,
and is essentially due to the fact that opinions are real-valued. In this case,
in fact, the probability for a "damaged" opinion to recover its value
in the unperturbed configuration is zero; on the other
hand, to perturb the opinion of an agent it suffices that one of its compatible 
neighbours be affected, and the probability of having a compatible "disturbed"
neighbour increases with the confidence bound $\epsilon$. 
The only circumstance which can stop the propagation of the damage is when 
the perturbed agents are not compatible with any of their neighbours.  
The considerations above allow us to understand why
the critical threshold $\epsilon_s=1/2$ found in \cite{Fortunatodam}, above which
damage spreads to all agents of the system, coincides with the threshold for
complete consensus of the model, as in this case all agents share the same
opinion and so they are all compatible with each other, which means that any agent
was affected by each of its neighbours at some stage.
Another interesting result of \cite{Fortunatodam} is the fact that the two
critical thresholds which separate the "damage" phases in the $\epsilon$-space
do not seem to depend on the degree $d_0$ of the first node affected by the shock, although
the Hamming distance at a given $\epsilon$ increases with $d_0$. This means
that it is irrelevant whether 
the shock initially affected somebody who has many social contacts or somebody 
who is instead poorly connected: if damage spreads in one case, it will 
do in the other too.

\begin{figure}
\centering
\includegraphics[angle=-90, scale=0.45]{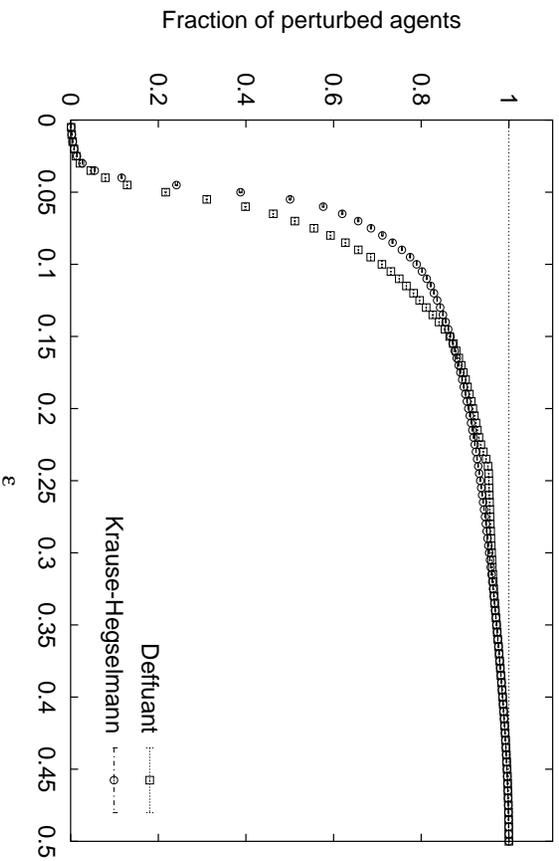}
\caption{Fraction of perturbed agents in the final configuration 
as a function of $\epsilon$
for the D and the KH model on a Barab{\'a}si-Albert network.}
\label{figd1}
\end{figure}

It is important to study as well how damage spreads under the D opinion
dynamics. The hope is to be able to identify common features 
which would allow to characterize the spreading process independently of the
specific consensus model adopted.
In section \ref{sec:4} we stressed the analogies between the KH and the D model, so we
expected to find similar results.
For the D model we need to fix one more parameter
to determine the dynamics, the convergence parameter $\mu$. 
The value of $\mu$ affects exclusively the time 
needed to reach the final configuration, so it has no influence on our results:
we set $\mu=0.3$.
Fig. \ref{figd1} shows how the Hamming distance varies with the confidence bound
$\epsilon$ for the D model on a BA network. The total number of agents is
$1000$. We remark that 
the damage is here calculated when the two replicas of the system attained their final
stable configurations.
We have also plotted the corresponding curve for the KH model, as obtained in 
\cite{Fortunatodam}. The two curves are quite similar, as we expected, and the
thresholds for the damage spreading transition are very close to each other.
Again, for $\epsilon>\epsilon_s=1/2$, all agents will be affected by the 
original perturbation.

As we explained in the introduction, our main aim is to attempt a 
spatial characterization of the damage spreading process, which 
would be impossible on a BA network. This is why from now on
we shall focus on the lattice topology. 
Here we start by changing the opinion variable of the agent lying on
the center site of
the lattice; if the lattice side $L$ is even, as in our case, 
the center of the
lattice is not a site, but the center of a plaquette, so we "shocked" one of the four
agents of the central plaquette. We refer to the initially shocked agent
as to the origin. 
We will address the following issues:

\begin{itemize}
\item{How far from the origin can the perturbation go?}
\item{What is the probability for an agent at some distance from the origin 
to be itself affected?} 
\item{How does this probability $p(d,t)$ vary with the distance $d$ and with
    the time $t$?}
\end{itemize}

To discuss the first issue, we need to calculate the {\it range} $r$
of the damage, i. e. the maximum of the distances from the origin of the agents reached
by the perturbation. The damage probability $p(d,t)$ is
the probability that, at time $t$, a randomly chosen agent at distance $d$ from
the origin changed its mind, due to the initial shock. Here the time is represented as usual
by the succession of opinion configurations created by the dynamics.
The time unit we adopted is 
one sweep over all agents of the system.
To calculate $p(d,t)$ we proceed as follows:
after $t$ iterations of the algorithm, we   
select all sites which are at the distance $d$ from the origin and which 
lie on-axis with respect to the origin, as in
the scheme below

\begin{center}
\includegraphics[angle=90, scale=0.45]{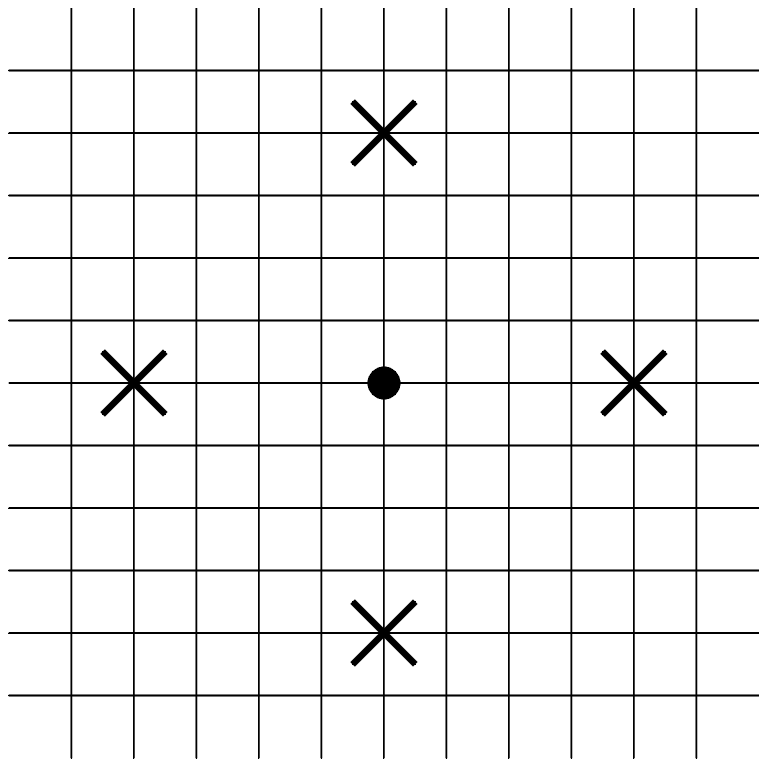}
\end{center}

\noindent where the black dot in the middle represents the origin and
the crosses mark the agents to be monitored. 
The damage probability 
is simply given by the fraction of these agents whose
opinions differ from those of their counterparts in the unperturbed
configuration (e. g. if two of the four agents changed their mind,
the probability is $2/4=1/2$). 
Note that by construction $d$ must be a multiple of the lattice spacing (in our
illustrated example $d=4$).
At variance with the evaluation of the damage range $r$, where we review all lattice
sites, for the damage probability
we neglected the off-axis sites because the
lattice is not isotropic and the corresponding data would be affected by strong
finite size effects due to the lack of rotational symmetry. 
To derive $p(d,t)$
only from four sites is of course difficult and we need to average
over many samples
for the data to have statistical meaning;
we found that a number of samples of the order of $10^3$ is enough to obtain stable results.
We calculated $p(d,t)$ for all distances from the center to the edges of the 
lattice and for all intermediate states of the system from the initial
random configuration to the final stable state.

\begin{figure}
\centering
\includegraphics[angle=-90, scale=0.45]{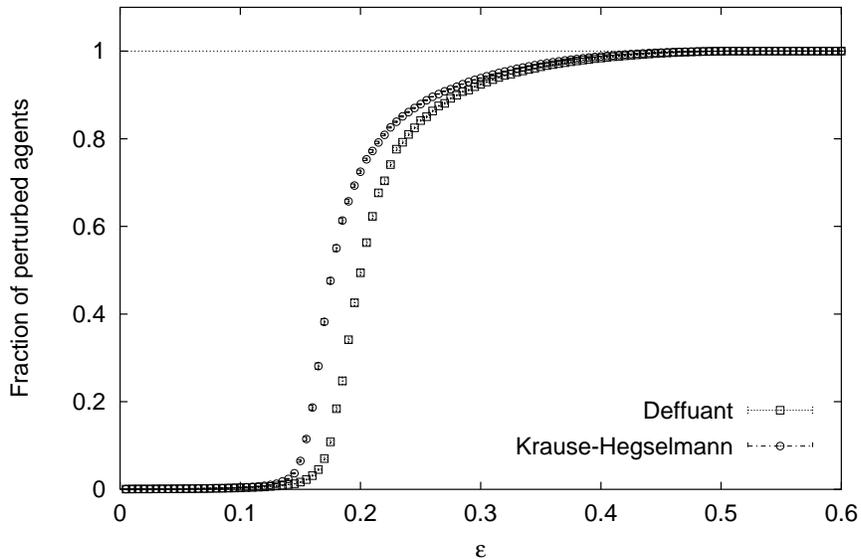}
\caption{As Fig. \ref{figd1}, but for 
agents sitting on the sites of a square lattice.}
\label{figd2}
\end{figure}
\begin{figure}
\centering
\includegraphics[angle=-90, scale=0.45]{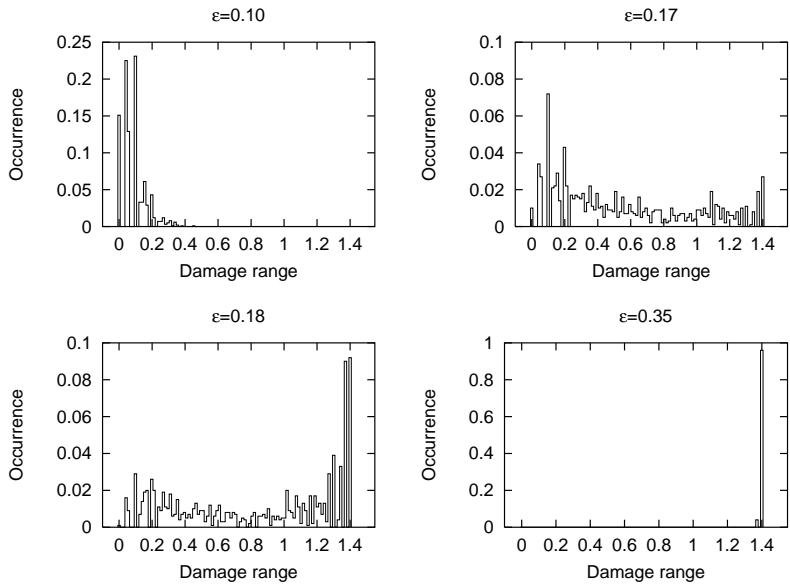}
\caption{D model, continuous opinions. Histograms of the damage range corresponding to 
four values of $\epsilon$; the lattice size is $40^2$.}
\label{figd3}
\end{figure}

We will present mostly results relative to 
the D model. The corresponding analysis for the KH model leads to  
essentially the same results.
For the purpose of comparison with Fig. \ref{figd1},
we plot in Fig. \ref{figd2}  
the Hamming distance as a function of $\epsilon$, for the D and the KH model.
The curves refer to a lattice with $40^2$ agents: 
the two patterns are again alike. The damage spreading thresholds are  
close, but they lie quite a bit higher than the corresponding values relative to the BA
network. This is basically due to the fact that in a BA network
each vertex lies just a few steps away from any other vertex (small world property),
and this makes spreading processes much easier and faster. Indeed, in the 
damage spreading phase, the time needed for the perturbation to 
invade the system is much longer 
for the lattice than for the BA network. 

\begin{figure}
\centering
\includegraphics[angle=-90, scale=0.45]{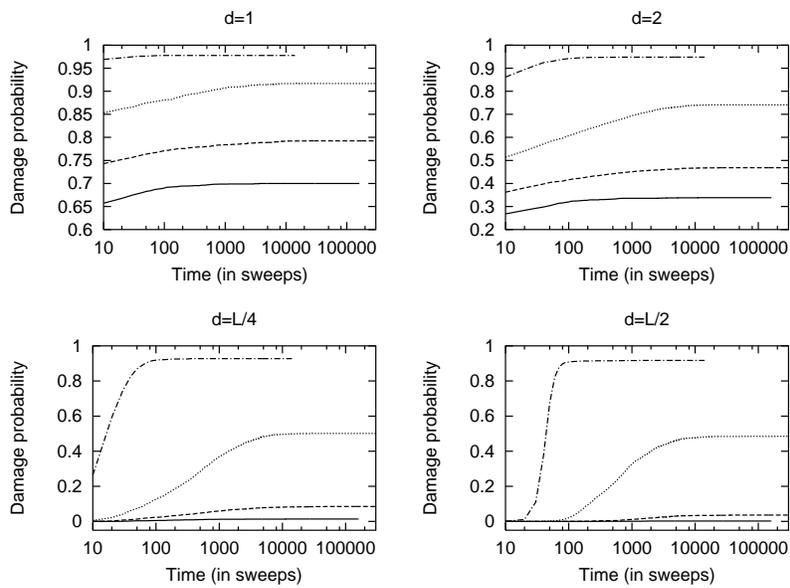}
\caption{D model, continuous opinions. Time evolution of the 
damage probability. Each frame refers to a fixed
distance $d$ from the origin, the curves are relative to different
values of $\epsilon$; the lattice size is $40^2$.}
\label{figd4}
\end{figure}

Since the amount of the damage is a function of $\epsilon$, 
the range $r$ of the damage is also a function of $\epsilon$.
It is interesting to analyze the histograms of the values of $r$ 
for different values of the confidence bound. In Fig. \ref{figd3} we show four
such histograms, corresponding to $\epsilon=0.10, 0.17, 0.18, 0.35$.
Note that the values of $r$ reported on the $x$-axis are expressed in units of
$L/2$ (half of the lattice side), which is
the distance of the center site from 
the edges of the lattice; 
since the farthest points from the origin 
are the four vertices of the square, the maximal possible value of
$r$ is $L\sqrt{2}/2$ (which corresponds to $\sqrt{2}\sim 1.414$ in the figure).
In the top left frame ($\epsilon=0.10$), damage does not spread and in fact 
the histogram is concentrated only at low values of $r$. In the 
other two frames, instead, we are near the threshold for damage spreading,
and we see that the damage often reaches the edge of the lattice ($r=1$ in the plot)
and even the farthest vertices ($r=\sqrt{2}$). The step 
from $\epsilon=0.17$ to $\epsilon=0.18$, in spite of the little difference in
the value of the confidence bound, is quite dramatic and signals the phase
transition: in the first case (top right) it is more likely to have short ranges
than long ones, in the other (bottom left) we have exactly the
opposite. In the last frame, the range is almost always maximal; looking at
Fig. \ref{figd2}, we can see that for $\epsilon=0.35$ more than $90\%$ of the 
agents are disturbed, so it is very likely that the perturbation reaches one of
the four vertices of the square.

\begin{figure}
\centering
\includegraphics[angle=-90, scale=0.45]{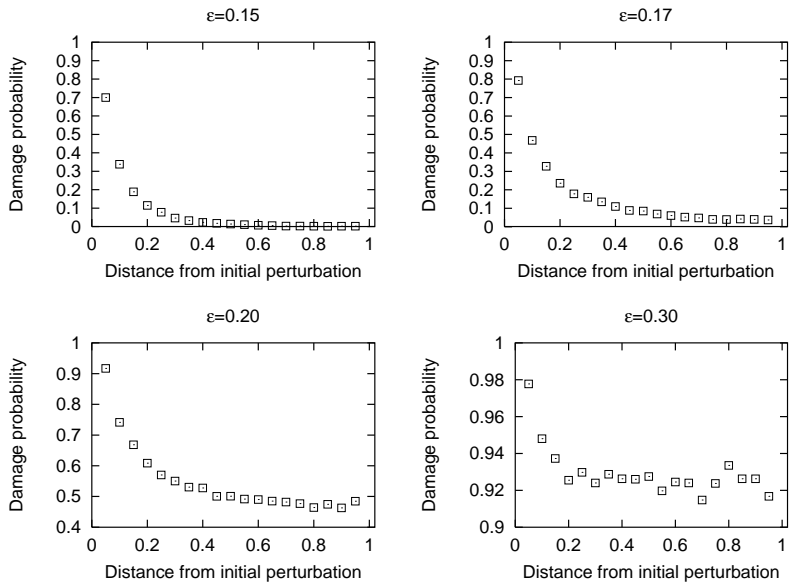}
\caption{D model, continuous opinions. 
Dependence of the damage probability on the distance $d$
  from the origin, when the system has reached the final stable configuration; 
the lattice size is $40^2$.}
\label{figd5}
\end{figure}

The study of the damage probability $p(d,t)$ is more involved, as it is a
function of two variables, the distance $d$ and the time $t$. 
A good working strategy is to analyze separately the dependence of $p(d,t)$ on the two
variables. We can fix the distance to some value $d_0$ and study how the 
damage probability at $d_0$ varies with time. We can also  
fix the time to $t_0$ and study how the probability at time $t_0$ 
varies with the distance
from the origin. On top of that, we should not forget the dependence on
$\epsilon$, which determines the "damage" state of the system.

In Fig. \ref{figd4} we explicitely plot the time dependence of the damage
probability at four different distances from the origin, $d=1, 2, L/4, L/2$.
In each frame, we have drawn four curves, corresponding (from bottom to top) 
to $\epsilon=0.15, 0.17, 0.20, 0.30$. We remark that the probability is the higher
the larger $\epsilon$, since this corresponds to a larger number of affected 
agents. All curves increase with time, which shows that
the damage does not heal, and they reach a plateau long before the system 
attains the final opinion configuration.
Note the rapid rise of the probability at the two largest 
distances ($L/4$ and $L/2$), for the two values of $\epsilon$ which fall in the
damage spreading phase ($\epsilon=0.20, 0.30$). 

Fig. \ref{figd5} shows how the damage probability varies with the distance from
the origin, at the end of the time evolution of the system. The values of the
distance on the $x$-axis are renormalized to the maximal distance on-axis
from the origin, $L/2$, as in Fig. \ref{figd3}.
We have again four frames, one for each of the four values of $\epsilon$ we have
considered in Fig. \ref{figd4}. 
We notice that 
for $\epsilon=0.15$, which is slightly below the threshold, the damage probability 
at the edge (top left) is zero, whereas for $\epsilon=0.17$, which is near the
threshold, it is small but nonzero (top right) and it is about $1/2$ for $\epsilon=0.20$
(bottom left). We tried to fit the curves with simple functions of the
exponential type. We found that 
the decrease with the distance 
is stronger than exponential: for low $\epsilon$, $p(d,t)$ (at fixed $t$) 
is well approximated by $a\,\exp(-bd)/d$.

We remind that we have chosen to introduce the shock in the system just at the
beginning of the evolution. If one instead would perturb the system some time
later, the amount of the damage and the corresponding probabilities would
decrease; however, the results of the analysis would be qualitatively the same.

\subsection{Discrete Opinions}
\label{sec:8}

There is essentially one reason which justifies 
the use of real-valued opinions:
the opinions of any two individuals are never exactly the same, although
they can be arbitrarily close. This is what commonly happens in society, where 
no two persons have exactly the same idea or judgement about any issue.
In fact, our opinion about somebody or a special event can fall anywhere between the
two edges "very bad" and "very good", something like the spectrum of visible
light, where one can pass smoothly from red to violet.

\begin{figure}
\centering
\includegraphics[angle=-90, scale=0.45]{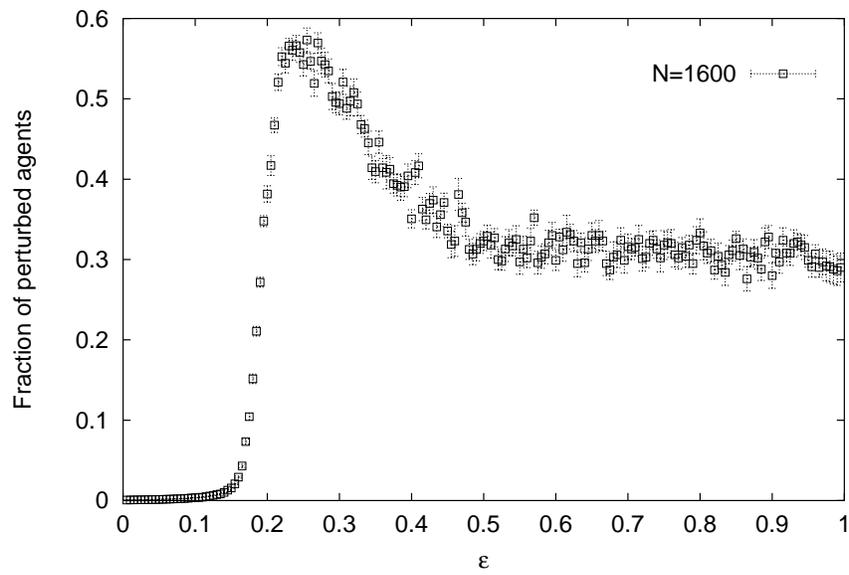}
\caption{D model, integer opinions. 
Fraction of perturbed agents in the final configuration 
as a function of $\epsilon$ for agents sitting on the sites of a square lattice.}
\label{figd6}
\end{figure}

On the other hand, for all practical purposes, this
continuous spectrum of possible choices can be divided in a finite number of
"bands" or
"channels", where each channel represents groups of close opinions. 
This is actually what teachers do when they evaluate
the essays of their students with marks, which are usually integers.
Also electors have to choose among a finite number of parties/candidates.
Finally, for the case we are mostly interested in, i. e. the reaction of people
to extreme events, the only possible quantitative investigation for sociologists
consists in making polls, 
in which the interviewed persons have to choose between a few options.

These examples show that it is more realistic to use integers rather 
than real numbers for the opinion variables of consensus models. 
Here we will repeat the damage spreading analysis of the previous section for 
the D model with integer opinions on a square lattice. 
We will see that the results are quite different from those we found before,
due to the phenomenon of {\it damage healing}.

To start with, we must
fix the total number $Q$ of possible opinions/choices. Since we performed
simulations for systems with few thousands agents, we decided to 
allow for a number of choices of the same order of magnitude, therefore we set  
$Q=1000$. The confidence bound must be an integer
$\ell$, but for consistency with the 
notation we have adopted so far, we will still use a real $\epsilon$, again
between $0$ and $1$, so that $\ell$ is the closest integer to $\epsilon\,Q$.

\begin{figure}
\centering
\includegraphics[angle=-90, scale=0.45]{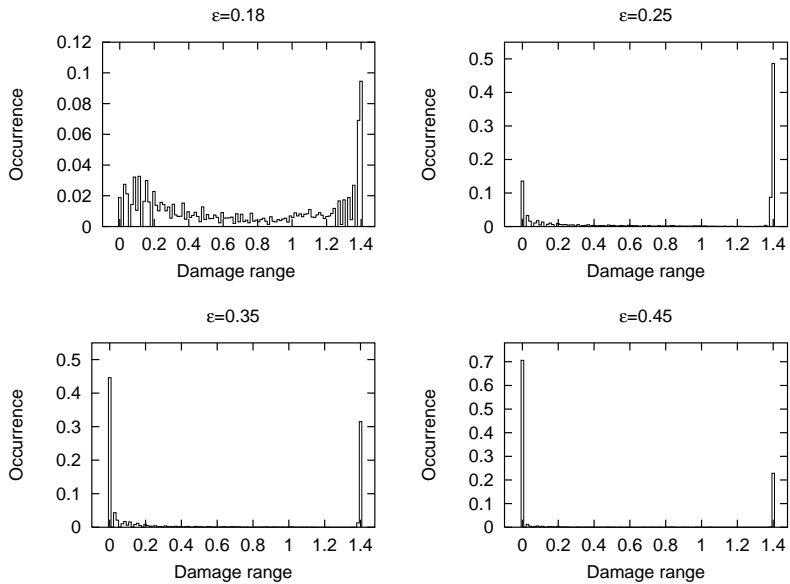}
\caption{D model, integer opinions. 
Histograms of the damage range corresponding to 
four values of $\epsilon$; the lattice size is $50^2$.}
\label{figd7}
\end{figure}

In Fig. \ref{figd6} we show the variation of the Hamming distance with the
confidence bound $\epsilon$, for a lattice with $40^2$ sites.
We immediately notice the difference with the analogous Fig. \ref{figd2} for
continuous opinions: after the rapid variation at threshold, 
the fraction of damaged sites reaches a peak, then it decreases and finally it 
forms a plateau at large $\epsilon$. Going from real to integer opinions we have 
no more total damage, i. e. the perturbation can affect at most some fraction $f<1$
of the total population (here $f\sim\,0.6$), but it has no chance to affect all agents.
If we increase the number of agents $N$ but we keep $Q$ fixed to the same value,
the height of the final plateau decreases, going to zero when $N/Q\rightarrow\infty$.

\begin{figure}
\centering
\includegraphics[angle=-90, scale=0.45]{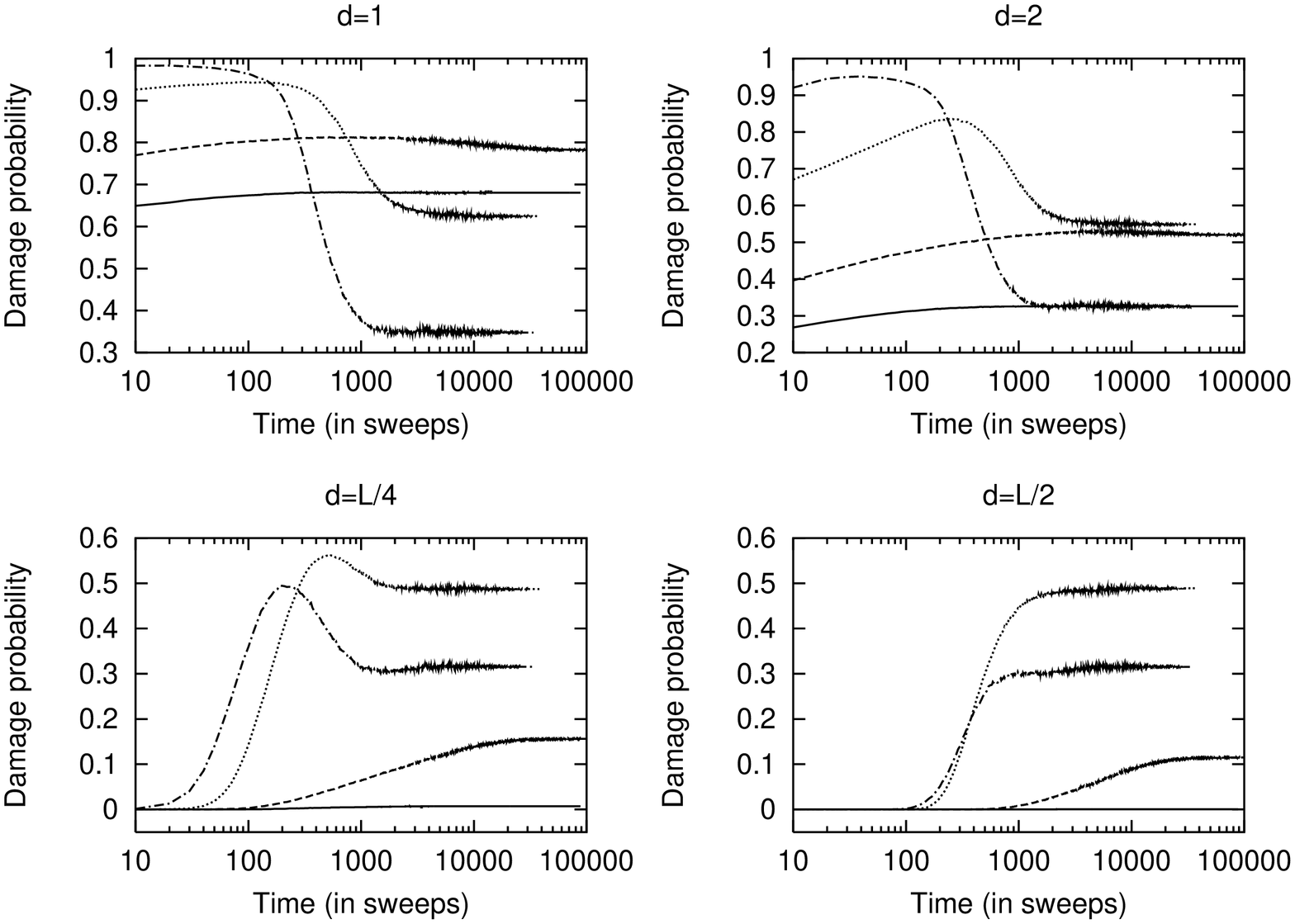}
\caption{D model, integer opinions. Time evolution of the 
damage probability. Each frame refers to a fixed
distance $d$ from the origin, the curves are relative to different
values of $\epsilon$; the lattice size is $50^2$.}
\label{figd8} 
\end{figure}

Why does this happen? Taking a look at Fig. \ref{figd7} helps to clarify the situation.
Here we see the histograms of the damage range for $\epsilon=0.18, 0.25, 0.35, 0.45$.
If we compare the frame relative to $\epsilon=0.18$ (top left) with its counterpart for
continuous opinions (Fig. \ref{figd3}, bottom left), we see that 
they are basically the same. We are close to the transition so there is some
finite
probability for the damage to reach the edges and even the vertices of the square. 
We notice that the histogram is continuous, in the sense that any 
value of the range between the two extremes is possible.
If we now look at the other three frames, the situation is very different:
the range can be either very short or very long. In particular, when $\epsilon$
is very large (bottom right), the range is zero or maximal. That means
that either the damage heals, or it spreads to all agents. 
In fact, for large $\epsilon$ ($>1/2$), there is complete 
consensus in the final configuration
(see section \ref{sec:3}), so all agents will end up with the same opinion.
The question is then whether the final opinion in the perturbed 
configuration coincides or not with that of the unperturbed configuration; in
the first case we have no damage, in the second total damage.

Now, real-valued opinions can be modified by arbitrarily small amounts, and that
would still correspond to damage. On the contrary, the variations of integer opinions 
are discontinuous steps, and the latter are much more unlikely to occur.
In this way, it is virtually impossible for a single agent
to trigger a "jump" of the final opinion of all agents of the system
to a different value. So, for large $\epsilon$ and many agents, the original
perturbation will be healed by the dynamics\footnote{The non-vanishing
probability for total damage in Fig. \ref{figd7} is a finite size effect,
as the total number $Q$ of opinions is about the same as the population $N$.} 
(no damage), whereas for 
continuous opinions even a small shock manages to shift a little bit the final
opinion of the community (total damage).

\begin{figure}
\centering
\includegraphics[angle=-90, scale=0.45]{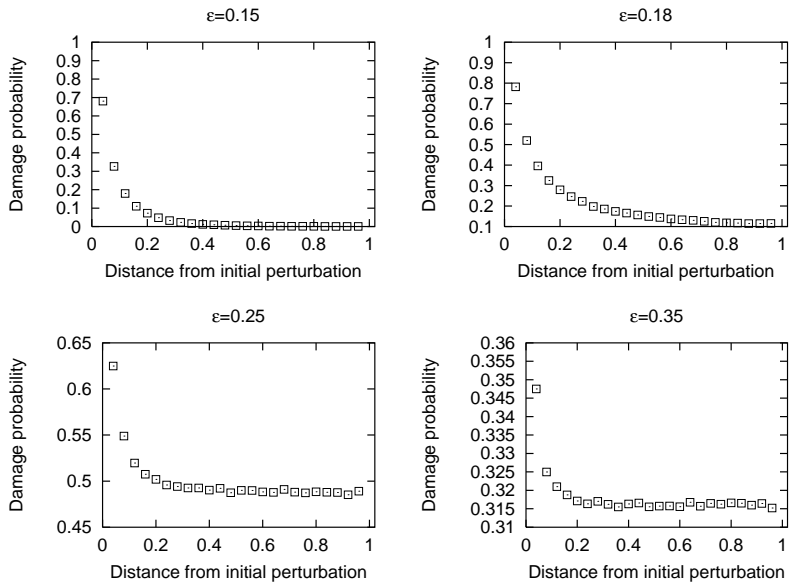}
\caption{D model, integer opinions. 
Dependence of the damage probability on the distance $d$
  from the origin, when the system has reached the final stable configuration; 
the lattice size is $50^2$.}
\label{figd9}
\end{figure}

The presence of damage healing is also clearly visible in 
Fig. \ref{figd8}, which is the counterpart of Fig. \ref{figd4} for integer opinions.
The four curves of each frame refer to $\epsilon=0.15$ (continuous), $0.18$
(dashed), $0.25$ (dotted) and $0.35$ (dot-dashed). The damage probability is no longer
monotonically increasing as in Fig. \ref{figd4}, but it displays various patterns,
depending on the confidence bound and the distance from the origin.
In particular, observe the behaviour of the 
curve for $\epsilon=0.35$ and $d=1$ (top left frame, dot-dashed line): here the
probability is initially close to $1$, because we are examining a neighbour   
of the shocked agent, but after few iterations it 
falls to about $0.3$, due to healing. We also note the curious shape of the 
two upper curves for $d=L/4$, which recalls the pattern of the Hamming distance
with $\epsilon$ of Fig. \ref{figd6}: the damage probability rapidly rises to a maximum
and then it decreases to an approximately constant value. 

Fig. \ref{figd9} shows the dependence of the
damage probability on the distance in the final opinion configuration,
for $\epsilon=0.15, 0.18, 0.25, 0.35$.
The curves look similar as those of Fig. \ref{figd5}. Again, the
damage probability decreases faster than exponentially.

\begin{figure}[hbt]
\begin{center}
\includegraphics[angle=-90,scale=0.45]{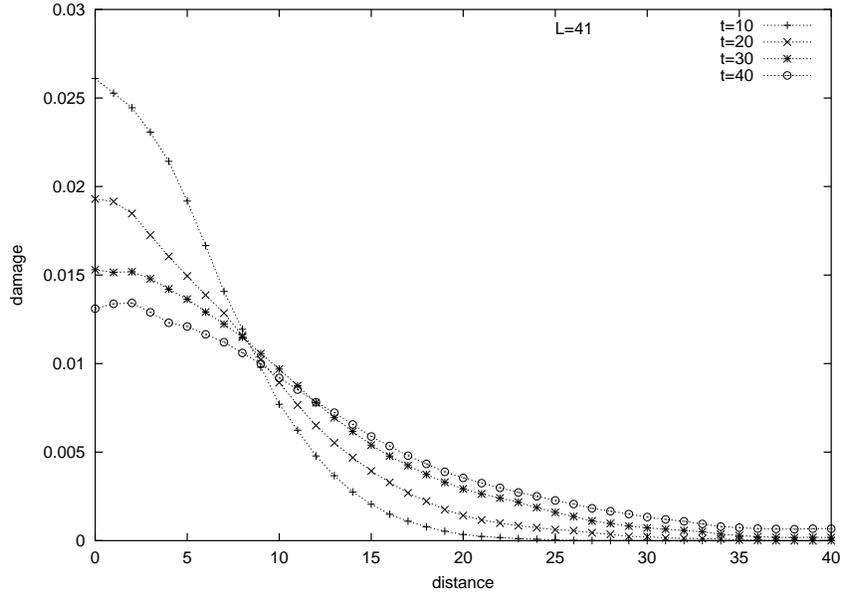}
\end{center}
\caption{\label{Klietsch} S model, two opinions.
Dependence of the damage probability on the distance $d$
  from the origin, for various times on a $41 \times 41$ lattice.}
\end{figure}

We conclude with some new results 
on damage spreading for the S model with two opinions 
on a square lattice \cite{Klietsch}, which complement the analyses
of \cite{Bernardes,Roehner}. 
Fig. \ref{Klietsch} shows the
damage probability as a function of distance at various times. We see that
the values of the probability are quite low; in fact, the system always evolves
towards consensus, so the damage will heal on the long run, as it happens
in the D and KH models (with discrete opinions)
when the confidence bound $\epsilon$ is above the threshold for
complete consensus.

\begin{figure}[hbt]
\begin{center}
\includegraphics[angle=-90,scale=0.45]{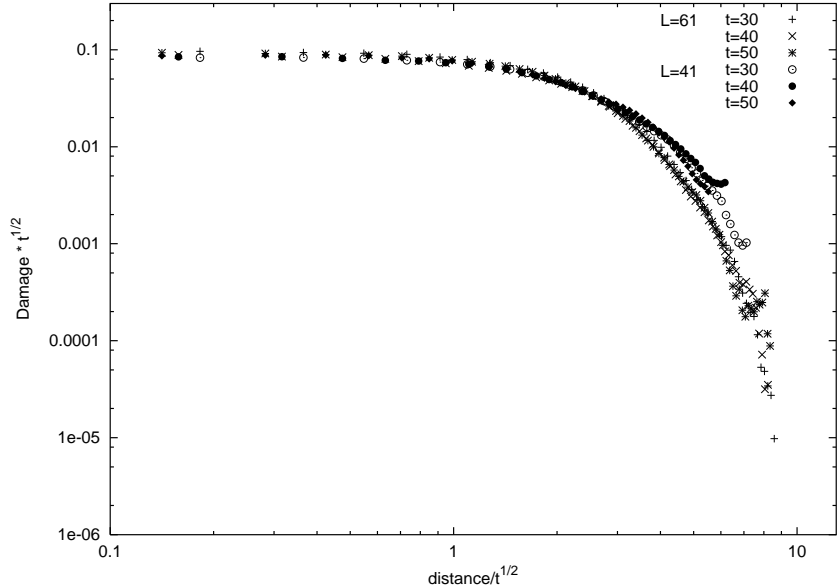}
\end{center}
\caption{\label{Klietsch1} S model, two opinions. 
Rescaling of the damage probability curves of Fig. \ref{Klietsch}, also for
larger lattices.
Here we plot the damage probability times $\sqrt{t}$ versus
$d/\sqrt t$ ($t$ is the time). For 
$t>20$, the curves for different times roughly overlap.}
\end{figure}

If damage would spread like in a diffusion process, the distance covered 
by the propagation of the perturbation would
scale as the square-root of the time $t$, and the probability to damage a site
at distance $d$ would follow for long times a scaling function $f(d/\sqrt
t)$. Fig. \ref{Klietsch1} shows that for $t \gg 1$ this seems indeed to be the case,
even though damage spreading is not a random diffusion process.

Applications of these techniques to the case of several different
themes on which people may have an opinion will be given elsewhere
\cite{Jacob}.

\section{Discussion}
\label{sec:9}

The three main models D, KH and S discussed follow different rules but give
similar results: They end up in a final state where no opinion changes anymore.
Depending on the confidence interval $\epsilon$ for continuous 
opinions, or $\ell$ for discrete opinions, this final state contains one opinion
(consensus), two (polarization) or three and more (fragmentation). In the 
discrete case with $Q$ different opinions, there is a maximum $Q$ (2 for D, 3
for S, 7 for KH) for which a consensus usually is found. These numbers may
correspond to the maximum number of political parties which may form a 
stable coalition government. The three rules differ
in that S describes missionaries who don't care about the previous opinions 
of those whom they want to convince; KH describes opportunists who follow the
average opinion of their discussion partners; and D describes negotiators 
who slowly move closer to the opinion of their discussion partner.  
Election results were successfully simulated by the model of S but not by 
that of D and KH, perhaps simply because nobody tried it yet with D and KH. 

The reaction of people to extreme events was investigated 
by performing a damage spreading analysis on the three 
consensus models we have introduced. The extreme
event induces a change of opinion in one (or a few) 
agent(s), the dynamics propagates the
shock to other agents. We represented the social 
relationships between people with a
square lattice and a scale-free network 
{\'a} la Barab{\'a}si-Albert. In both
cases we found that there is quite a wide range 
of values of the confidence interval $\epsilon$ 
(or $\ell$) for which the original shock 
influences the opinions of a non-negligible fraction 
of the community.  For very tolerant people and 
continuous opinions, the whole community will be 
affected by the event on the long run. By using 
integer-valued opinions, instead, we found that 
the perturbation cannot affect more than a maximal 
fraction of the population (it can be sizeable, 
though). On the lattice we could as well study  
how the influence of the extreme event on the 
opinions varies with the distance in time and space 
from the event. The damage probability at a fixed 
distance from the original shock varies very rapidly 
with time; it increases up to a plateau for 
continuous opinions, it follows more involved 
patterns for integer opinions.  
Our analysis also shows that the effect of 
the perturbation falls faster than exponentially
with the distance from the place where the event took place.

What have we achieved with these simulations? We did {\em not} find a way
to predict earthquakes or floods, nor did we propose a method how to
convince people to judge these dangers objectively, instead of being
overly influenced by events close in time and space, and of forgetting
the lessons from distant catastrophes which happened long ago. Our simulations
give quantitative data for these space-time correlations of opinions and 
extreme events. Once sociology delivered quality data on real people and
their opinions \cite{Pohl}, one can compare these results with the simulations
and modify if needed the simulations until they give a realistic description.
Only then can the simulations be used to predict how danger perception will 
develop in space and time.

We thank J.S. S\'a Martins for a critical reading of the manuscript.
SF gratefully acknowledges the financial support of the DFG
Forschergruppe under grant FOR 339/2-1.

\end{document}